\documentclass{JHEP3}

\usepackage{Pformel,Pmeta}
\usepackage{amsmath,amsfonts,amsthm,times}

\title{Seiberg-Witten equations and non-commutative spectral curves in Liouville theory}

\author{Leonid Chekhov$^{1,2}$, Bertrand Eynard$^3$ and Sylvain Ribault$^{3,4}$
\\
\!\!\!$^1$\! Department of Theoretical Physics, Steklov Mathematical Institute
\\
Moscow, 119991 Russia
\\
\!\!\!$^2$\! School of Mathematics, Loughborough University 
\\
LE11 3TU, Leicestershire, UK
\\
\!\!\!$^3$\! Institut de Physique Th\'eorique, IPhT
\\
CNRS, URA 2306
\\
F-91191 Gif-sur-Yvette, France
\\
\!\!\!$^4$\! Laboratoire Charles Coulomb UMR 5221 CNRS-UM2
\\
Universit\'e Montpellier 2
\\
Place Eug\`ene Bataillon, F-34095 Montpellier Cedex 5, France
\vspace{2mm}
 \\
 {\footnotesize \tt chekhov@mi.ras.ru, bertrand.eynard@cea.fr, sylvain.ribault@cea.fr }
}

\abstract{We propose that there exist generalized Seiberg-Witten equations in the Liouville conformal field theory, which allow the computation of correlation functions from the resolution of certain Ward identities. These identities involve a multivalued spin one chiral field, which is built from the energy-momentum tensor. We solve the Ward identities perturbatively in an expansion around the heavy asymptotic limit, and check that the first two terms of the Liouville three-point function agree with the known result of Dorn, Otto, Zamolodchikov and Zamolodchikov. 
We argue that such calculations can be interpreted in terms of the geometry of non-commutative spectral curves.
}

\makeatletter\let\default@color\current@color\makeatother 

\preprint{}

\begin{document}

\section{Introduction}

Among two-dimensional conformal field theories, Liouville theory may be viewed as the simplest non-trivial theory with a continuous spectrum. Unlike minimal models, which exist for discrete values of the central charge $c$, Liouville theory can be defined for any real value $c\geq 25$, and analytically continued to arbirtrary complex values of $c$. Like minimal models, Liouville theory is minimal, in the sense that it is completely determined by symmetry and consistency conditions, once the spectrum of primary fields is given. This spectrum is specified by the set of allowed conformal dimensions $\Delta \geq \frac{c-1}{24}$, with multiplicity one for each dimension. The Dorn-Otto-Zamolodchikov-Zamolodchikov formula for the three-point function \cite{do94,zz95} is the unique solution of the crossing symmetry equations for this spectrum \cite{tes95,tes03b}. And with the conformal bootstrap method \cite{bpz84}, it is in principle enough to know the three-point function for computing all correlation functions. 

The minimal nature of Liouville theory probably explains why this theory has many fundamental applications, in particular in two-dimensional gravity and exactly solvable models of string theory. In these applications, the conformal transformations which are symmetries of Liouville theory are respectively interpreted as space-time diffeomorphisms and worldsheet reparametrizations. 
More recently, applications to four-dimensional gauge theories were initiated by the work of Alday, Gaiotto and Tachikawa \cite{agt09}. The existence of the AGT relation between Liouville and gauge theories may seem surprising, because two-dimensional conformal transformations do not usually appear in gauge theories. But the gauge theories in question belong to a special class of supersymmetric theories, where such transformations do appear. 

Since Liouville theory is so fundamental, it is worthwhile to investigate the structure of the theory, and to look for alternative methods of computing its correlation functions. We are motivated by the following question: {\it are the correlation functions of Liouville theory encoded in the geometry of some spectral curves?} This is inspired in particular by the relations of Liouville theory with matrix models, whose observables are known to be encoded in certain algebraic curves. However, matrix model constructions cannot be directly relevant to Liouville theory, which has
no integer parameter corresponding to the integer size of the matrix. Only particular correlation functions in Liouville theory, where a continuous parameter takes an integer value, are related to matrix integrals. (See Appendix \ref{appmm} for more details.) However, the Topological Recursion formalism \cite{eo08}, which was originally developed in the context of matrix models in order to compute the observables from the geometry of the spectral curve, has a more general validity, and requires neither a matrix integral nor an integer parameter. 

In this article we will propose an alternative derivation of Liouville correlation functions, based on techniques inspired from Topological Recursion. The logic of the argument is:
\begin{enumerate}
 \item 
The basic principle of the relation of Liouville theory with matrix models (and of the AGT relation) is that each correlation function of Liouville theory can be interpreted as the partition function of a specific model,
\begin{align}
 Z = \la \prod_{i=1}^n V_{\al_i}(z_i,\bar{z}_i)\ra \ ,
\label{z}
\end{align}
so that the variables $\al_i$ (the momenta) and $z_i$ (the positions in the two-dimensional space) of our Liouville $n$-point function are interpreted as parameters of a model whose partition function is $Z$. 
\item
Using the holomorphic energy-momentum tensor $T$, we define a chiral (i.e. locally holomorphic) field $J$ by
\begin{align}
 T = -J^2 +Q \p J\ ,
\label{j}
\end{align}
where $Q$ is defined by $c=1+6Q^2$.
We introduce the correlation functions
\begin{align}
 \hat{W}_m(x_1,\cdots x_m) = \lla \prod_{j=1}^m J(x_j) \rra \ ,
\label{w}
\end{align}
where we define the double brackets by
\begin{align}
 \lla {\cal O} \rra = \frac{1}{Z} \la {\cal O} \prod_{i=1}^n V_{\alpha_i}(z_i,\bar{z}_i) \ra\ .
\label{oo}
\end{align}
 Then the correlation functions $\hat{W}_m$ obey Ward identities, which are analogous to the loop equations of matrix models.
\item
In order to derive the partition function $Z$ from the study of the correlation functions $\hat{W}_m$, we need to assume the validity of a Seiberg-Witten equation of the form 
\begin{align}
 \pp{\al_i} F = \int_{B_i} \hat{W}_1(x) dx \ ,
\label{sw}
\end{align}
where $B_i$ is some contour of integration, and the free energy $F$ is defined by
\begin{align}
 F= \log Z\ .
\end{align}
The Seiberg-Witten equation is inspired by matrix models and supersymmetric gauge theory, and it can be heuristically justified from the functional integral representation of Liouville correlation functions. We will treat it as a basic assumption, which will be justified if it leads to reasonable results. 
\item 
Finally, we can ask whether there exists a spectral curve $C$ where the $1$-cycle $B_i$, and the $1$-form $\hat{W}_1(x)dx$ live. However, this "exact" spectral curve is not very useful or accessible. As we will perform calculations in a perturbative expansion of Liouville theory, we will rather discuss the "perturbative" curve which is associated to that particular expansion, and on which the leading term $\hat{W}_1^{(0)}(x)dx$ of $\hat{W}_1(x)dx$ lives. More precisely, we will consider the heavy asymptotic limit 
\begin{align}
 Q \rar \infty \scs \eta_i=\frac{\al_i}{Q} \ \ {\rm fixed}\ ,
\label{ci}
\end{align}
and the corresponding perturbative expansion. This leads to the non-commutative curve
\begin{align}
 y^2 + t^{(0)}(x) = 0 \scs [y,x]=1 \ ,
\label{yt}
\end{align}
where $t^{(0)}(x) = \underset{Q\rar \infty}{\lim} Q^2\lla T(x) \rra$ is a rational function of $x$ with second-order poles at $z_i$. 
Using the identification $y=\pp{x}$, the equation of our curve can be viewed as a differential equation, namely
the Fuchsian form of the Liouville equation with $n$ sources. 
\end{enumerate}
Our concrete aim is to rederive the DOZZ formula for the Liouville three-point function using these ideas. We will partly fulfill this aim, by rederiving the first two terms of the expansion of that formula around the heavy asymptotic limit. This will be done using the formalism of \cite{cem09} for correlation functions associated to non-commutative spectral curves. (However, no previous knowledge of that article or of Topological Recursion is assumed.)

\section{The spin one chiral field in Liouville theory}

In this section, we study the chiral field $J$ defined by eq. (\ref{j}). Some readers may object firstly that the solution of the quadratic equation (\ref{j}) cannot be a unique, well-defined chiral field $J$, and secondly that if it was, it would be a $\hat{u}(1)$ current, so that we would be studying a free boson instead of Liouville theory. The answer to both objections is that $J$ should be considered as a multivalued chiral field, which is  locally holomorphic and has not only poles like the energy-momentum tensor $T$, but also branch cuts. Due to these branch cuts, some of the Ward identities of the free boson theory fail, and $J$ does not encode a $\hat{u}(1)$ symmetry. The relation between the energy-momentum tensor $T$ and the chiral field $J$ is actually nothing but the operator version of the usual construction of the Virasoro algebra in terms of the Heisenberg algebra.

\subsection{Extended chiral algebra}

The symmetry algebra of Liouville theory is the Virasoro algebra of local conformal transformations. This algebra is encoded in the operator product expansion (OPE) of the energy-momentum tensor $T$ with itself,
\begin{align}
 T(x) T(x') = \frac{\frac{c}{2}}{(x-x')^4} + \frac{2}{(x-x')^2} T(x') + \frac{1}{x-x'}\p T(x') + O(1)\ ,
\label{tt}
\end{align}
where $c$ is the central charge. The vertex operators $V_\al(z,\bar{z})$ are defined by their OPE with the energy-momentum tensor,
\begin{align}
 T(x) V_\al(z,\bar{z}) = \left(\frac{\Delta_\al}{(x-z)^2} + \frac{1}{x-z}\pp{z}\right)V_\al(z,\bar{z}) + O(1)\ ,
\label{tv}
\end{align}
where the conformal dimension $\Delta_\al$ associated to the momentum $\al$ is
\begin{align}
 \Delta_\al = \al(Q-\al) \ ,
\label{da}
\end{align}
where we recall the relation between the central charge $c$ and the background charge $Q$,
\begin{align}
 c = 1+6Q^2 \ .
\label{cq}
\end{align}
 Now in the free boson theory the symmetry algebra is the affine algebra $\hat{u}(1)$, which is encoded in the OPE of the chiral field $J$ with itself,
\begin{align}
 J(x)J(x') = \frac{-\frac12}{(x-x')^2} + J^2(x') + O(x-x') \ .
\label{jj}
\end{align}
The enveloping algebra of this $\hat{u}(1)$ algebra contains a Virasoro algebra with central charge $c$, whose energy-momentum tensor $T$ is constructed from $J$ by eq. (\ref{j}). We have written the $JJ$ OPE up to its first regular term $J^2(x')$, as this provides the definition for the $J^2$ term in eq. (\ref{j}).
The field $J$ is quasiprimary of conformal dimension one with respect to $T$,
\begin{align}
 T(x)J(x') = \frac{Q}{(x-x')^3} + \pp{x'} \frac{1}{x-x'} J(x') + O(1) \ . 
\label{tj}
\end{align}
The formal identity of our construction with the free boson theory shows that the set of OPEs (\ref{tt}), (\ref{jj}), (\ref{tj})  defines an associative algebra of operators. 

The crucial difference between the Liouville and free boson theories appears in the OPE of $J$ with a vertex operator $V_\al(z,\bar{z})$, which in the free boson theory reads
\begin{align}
 J(x) V_\al(z,\bar{z})\ \underset{\rm (free\ boson)}{=}\ \frac{-\al}{x-z} V_\al(z,\bar{z}) + O(1) \ . 
\label{jv}
\end{align}
This equation expresses the conservation of the momentum $\al$ which comes from the $\hat{u}(1)$ symmetry. This symmetry is absent in Liouville theory. Actually, vertex operators with momenta $\al$ and $Q-\al$ have the same conformal dimension $\Delta_\al$ (\ref{da}), so that they must be identified, as a consequence of the assumption that each representation of the Virasoro algebra appears with multiplicity one in the spectrum of Liouville theory:
\begin{align}
 V_\al(z,\bar{z}) = R(\al) V_{Q-\al}(z,\bar{z})\ ,
\label{vrv}
\end{align}
where $R(\al)$ is a reflection coefficient \cite{zz95}. This reflection equation is clearly incompatible with the OPE (\ref{jv}) as it stands. Notice however that we can make sense of that OPE in the context of the idea of a multivalued field $J$. Looking for a $J(x) V_{\al_i}(z_i,\bar{z}_i)$ OPE which would be compatible with the known $T(x) V_{\al_i}(z_i,\bar{z}_i)$ OPE (\ref{tv}) given the relation (\ref{j}) between $T(x)$ and $J(x)$, we find the two solutions
\begin{align}
 \bla J(\overset{i+}{x}) V_{\al_i}(z_i,\bar{z}_i)= \frac{-\al_i}{x-z_i} V_{\al_i}(z_i,\bar{z}_i) + O(1)\ , 
\\ J(\overset{i-}{x}) V_{\al_i}(z_i,\bar{z}_i)= \frac{\al_i-Q}{x-z_i} V_{\al_i}(z_i,\bar{z}_i) + O(1) \ , 
\ela
\label{jvjv}
\end{align}
where we take $\pm$ to label sheets of the space where $J$ lives. Such sheets are not expected to be globally defined, and the labels $\pm$ in this OPE make sense only in the neighbourhood of $x=z_i$.

\subsection{Seiberg-Witten equations}

We now look for Seiberg-Witten equations involving correlation functions of the field $J$.
Since the field $J$ is multivalued, its correlation functions $\hat{W}_m$ (\ref{w}) and in particular $\hat{W}_1$ are multivalued. And the 
$JV$ operator product expansion (\ref{jvjv}) implies that in the neighbourhood of $x=z_i$ we have $\hat{W}_1(\overset{i+}{x}) = \frac{-\al_i}{x-z_i}+O(1)$, which leads to
\begin{align}
 \al_i = -\frac{1}{2\pi i}\oint_{z_i} \hat{W}_1(\overset{i+}{x})dx\ .
\end{align}
This equation is of the type of
$ \al_i = \int_{A_i} \hat{W}_1(x)dx$, which is usually associated to an equation of the type of eq. (\ref{sw}) provided  
the contour $A_i$ is dual to the contour $B_i$ of that equation. 
This suggests that eq. (\ref{sw}) may hold, provided the contour $B_i$ ends at $z_i$. Then $B_i$ must have another endpoint.
The choice of this other endpoint is problematic, and we will circumvent the problem by considering $\left(\pp{\al_i}-\pp{\al_j}\right)F$ instead of just $\pp{\al_i}F$. The natural Seiberg-Witten equation for this quantity is 
\begin{align}
 \left(\pp{\al_j}-\pp{\al_i}\right)F = \int_{z_i}^{z_j} \left(\hat{W}_1(\overset{i+}{x})+\hat{W}_1(\overset{j+}{x})\right)dx\ ,
\label{swp}
\end{align}
whose right-hand side is antisymmetric under $i\lrar j$ as it should be. 

It may seem that our choice of using the solution $\hat{W}_1(\overset{i+}{x})$ and not $\hat{W}_1(\overset{i-}{x})$ is arbitrary -- and incompatible with the reflection symmetry eq. (\ref{vrv}). We will later justify this choice under some assumptions (\ref{re}) on the values of parameters. And the results which we will obtain using our Seiberg-Witten equation will turn out to be compatible with reflection symmetry.

In Appendix \ref{apphsw}, we will prove generalizations of the Seiberg-Witten equations, which involve higher correlation functions $\hat{W}_m$ instead of $F$ and $\hat{W}_1$. 

\section{Ward identities}

The multivalued field $J$ is of course expected to have multivalued correlation functions. However, this issue can be ignored for the purpose of deriving Ward identities. The issue will matter only when we start solving these identities in Section \ref{subsecewi}.

\subsection{Non-perturbative Ward identities}

It is convenient to write identities not for the correlation functions $\hat{W}_m$ (\ref{w}), but for their connected components $W_m$ defined as
\begin{align}
 \hat{W}_1(x) &= W_1(x)\ ,
\\
\hat{W}_2(x_1,x_2) &= W_2(x_1,x_2)-\frac{\frac12}{(x_1-x_2)^2} + W_1(x_1)W_1(x_2)\ ,
\\
\hat{W}_3(x_1,x_2,x_3) &= W_3(x_1,x_2,x_3) + \sum_{j=1}^3 W_1(x_j)\left(W_2(x_k,x_\ell)-\frac{\frac12}{(x_k-x_\ell)^2}\right) + \prod_{j=1}^3 W_1(x_j) \ ,
\end{align}
which are examples of the general definition
\begin{align}
 \hat{W}_{|I|}(I) = \sum_{p\in P(I)} \prod_{J\in p} \left(W_{|J|}(J)-\delta_{|J|,2}\frac{\frac12}{(\Delta x(J))^2} \right) \ ,
\label{wii}
\end{align}
where $I$ is a finite set of variables of length $|I|$, $P(I)$ is the set of partitions of $I$ into nonempty subsets, 
and $\Delta x(J)$ is the difference of the two variables belonging to a set $J$ of size two. So, $\hat{W}_2(x,y)$ has a second-order pole at $x=y$ due to the $JJ$ OPE (\ref{jj}), but this pole is explicitly subtracted in the above definition, so that 
$W_2(x,y)$ is regular at $x=y$. Like $\hat{W}_m$, $W_m$ is symmetric under permutations of its arguments $(x_1,\cdots x_m)$.

The idea which leads to the Ward identities is to express a correlation function of the type of $\la T(x) \prod_{j=1}^m J(x_j) \prod_{i=1}^n V_{\al_i}(z_i)\ra$ in two different ways. On the one hand, the $x$-dependence can be fully determined by the axiom that $T(x)$ is meromorphic with poles only at $x_j$ and $z_i$, and that the behaviour of $T(x)$ near these poles is determined by the $TJ$ (\ref{tj}) and $TV$ (\ref{tv}) OPEs. (In particular, $T(x)$ is supposed to be regular at $x=\infty$ in the sense $T(x)\underset{x\rightarrow \infty}{=}o(1)$.)
On the other hand, the $T$ insertion can be written in terms of $J$ using the definition (\ref{j}) of $J$ (together with eq. (\ref{jj}) for the regularization of $J^2$).

In the case $m=0$, we thus want to express $\lla T(x) \rra$ in two different ways. On the one hand, meromorphy of $T(x)$ leads to 
\begin{align}
 \lla T(x) \rra = t(x) \ ,
\label{tet}
\end{align}
where the double brackets are defined in eq. (\ref{oo}), and
we define the function $t(x)$ as 
\begin{align}
 t(x) &= \sum_{i=1}^n \left(\frac{\Delta_{\al_i}}{(x-z_i)^2} + \frac{\beta_i}{x-z_i}\right)\ ,
\label{tx}
\\
\beta_i &= \pp{z_i} \log Z \ .
\label{be}
\end{align}
We will call $\beta_i$ the quantum accessory parameters. This is because the leading term of $\beta_i$ appears as an accessory parameter in the second-order differential equation (\ref{ptp}) (see for instance \cite{tz01}), which is 
the heavy asymptotic limit of the first Ward identity.
On the other hand, inserting the definition (\ref{j}) of $J$ into eq. (\ref{tet}) leads to the first Ward identity,
\begin{align}
 Q\pp{x} W_1(x) - W_1(x)^2 - W_2(x,x) = t(x)\ .
\label{wiz}
\end{align}
This is the $m=0$ case of the Ward identity, which is obtained an equality between two different expressions for $\lla T(x) \prod_{j=1}^m J(x_j) \rra$,  
\begin{multline}
 Q\pp{x} \hat{W}_{m+1}(I,x) -\underset{x'\rar x}{\lim}\left(\hat{W}_{m+2}(I,x,x')+\frac{\frac12}{(x-x')^2} \hat{W}_m(I)\right)
\\
= \left(t(x) + D_x +\sum_{j=1}^m\pp{x_j} \frac{1}{x-x_j}\right)\hat{W}_m(I) +Q\sum_{j=1}^m \frac{1}{(x-x_j)^3} \hat{W}_{m-1}(I-\{x_j\})\ ,
\label{qph}
\end{multline}
 where we denote $I=(x_1,x_2,\cdots x_m)$, and we introduce the differential operator
\begin{align}
 D_x = \sum_{i=1}^n \frac{1}{x-z_i}\pp{z_i}\ .
\label{dx}
\end{align}
Rewriting these Ward identities in terms of the connected correlation functions $W_m$ defined by eq. (\ref{wii}) yields
\begin{multline}
 Q\pp{x} W_{m+1}(I,x) -\sum_{J\subset I}W_{|J|+1}(J,x)W_{m-|J|+1}(I-J,x) - W_{m+2}(I,x,x)
\\
= \delta_{m,0}t(x) + D_xW_m(I)+\sum_{j=1}^m\pp{x_j}\frac{W_m(I)-W_m(I-\{x_j\},x)}{x-x_j}\ ,
\label{qpw}
\end{multline}
whose special case $m=0$ coincides with eq. (\ref{wiz}). Notice that for $m\neq 0$ each term of the sum over the subsets $J$ of $I$ appears twice, as the $J$ and $I-J$ terms coincide. 
The Ward identities are special cases of the loop equations of \cite{cem09} (see eq. (5.9) therein), so that we can use the formalism of that article when it comes to solving them. Our identities are more specific than the loop equations, which involve undetermined polynomial functions. This will allow us to go further in the resolution of Ward identities and calculation of free energies than can be done in the general setting of \cite{cem09}.

One may however wonder how this formalism can possibly help us compute the partition function $Z$. Using the Seiberg-Witten equations (\ref{sw}) we can deduce $Z$ from $W_1$, which can itself be obtained by solving eq. (\ref{wiz}) if we know $W_2$, whose calculation as the solution of eq. (\ref{qpw}) with $m=1$ requires the knowledge of $W_3$, and so on. On top of this, each Ward identity comes with its own integration constant. However, it will turn out that in the perturbative expansion of Section \ref{subsecewi}, computing $W_m$ up to the order $g$ involves only $W_1,W_2,\cdots W_{m+g}$. For example, eq. (\ref{wiz}) will lead to a closed equation for the leading term of $W_1$, as the $W_2$ term will be subleading. Moreover, the integration constants will be uniquely determined by simple analyticity conditions. 

In the case of the free boson theory, the field $J(x)$ in $\hat{W}_1(x)=\lla J(x)\rra$ is assumed to be meromorphic with poles only at $z_i$, and the behaviour of $J(x)$ near these poles is determined by 
the $JV$ (\ref{jv}) OPE. This leads to 
\begin{align}
 \hat{W}_1(x) \ \underset{\rm (free\ boson)}{=}\ -\sum_{i=1}^n \frac{\al_i}{x-z_i}\ . 
\label{wfb}
\end{align}
This can be generalized to
$\lla J(x)\prod_{j=1}^m J(x_j) \rra$, 
\begin{align}
 \hat{W}_{m+1}(I,x) \ \underset{\rm (free\ boson)}{=}\ \hat{W}_1(x)\hat{W}_m(I) 
-\sum_{j=1}^m\frac{\frac12}{(x-x_j)^2}\hat{W}_{m-1}(I-\{x_j\})\ .
\end{align}
This implies the vanishing of all higher connected correlation functions,
\begin{align}
 W_{m\geq 2} \ \underset{\rm (free\ boson)}{=}\  0\ .
\end{align}

\subsection{Global conformal symmetry, and the case $n=3$}

Let us analyze how correlation functions $W_m$ behave under global conformal transformations $z\rar \frac{az+b}{cz+d}$. 
Symmetry under such transformations constrains how $W_m$ depends on $(x_1,\cdots x_m)$ and $(z_1,\cdots z_n)$, and if $n\leq 3$ these constraints can be interpreted as completely determining the dependence on $(z_1,\cdots z_n)$.

The consequences of global conformal symmetry can be neatly encoded in the behaviour of the chiral fields $T(x)$ and $J(x)$ as $x\rar \infty$, 
\begin{align}
 T(x)=O\left(\frac{1}{x^4}\right) \scs J(x) = -\frac{Q}{x} + O\left(\frac{1}{x^2}\right)\ ,
\label{tiqi}
\end{align}
which follow from the $TT$ and $TJ$ OPEs, and the assumption that the energy-momentum tensor $T$ generates the conformal transformations.
This asymptotic behaviour must also apply to correlation functions of $T$ and $J$. For instance, in the free boson theory, the study of the asymptotic behaviour of $\hat{W}_1(x)$ (\ref{wfb}) as $x\rar\infty$ leads to the conservation of momentum, $\sum_{i=1}^n \al_i=Q$. 

The Ward identity for $T$ (\ref{tet}) implies $t(x)=O\left(\frac{1}{x^4}\right)$, which constrains the quantum accessory parameters $\beta_i$ of eq. (\ref{tx}). In the case $n=3$ the accessory parameters are completely determined by such constraints, and we have 
\begin{align}
 t(x)\ \underset{(n=3)}{=}\ \frac{1}{\varphi(x)} \sum_{i=1}^3 \Delta_{\al_i} \frac{\varphi'(z_i)}{x-z_i} \ ,
\label{tf}
\end{align}
where we introduce the function
\begin{align}
 \varphi(x)= \prod_{i=1}^3 (x-z_i)\ .
\label{vf}
\end{align}
More generally, consider the Ward identity
(\ref{qph}), which is a relation between two expressions for $\lla T(x)\prod_{j=1}^m J(x_j)\rra$. In particular, 
 its right-hand side must behave as $O\left(\frac{1}{x^4}\right)$ as $x\rar \infty$. 
Requiring the vanishing of the $O(\frac{1}{x}),O(\frac{1}{x^2})$ and $O(\frac{1}{x^3})$ terms leads to three partial differential equations for $\hat{W}_m$. This leads to constraints on the dependence of $W_m$ on $(x_1,\cdots x_m)$, whose general solutions are 
\begin{align}
W_1(x) &= -\frac{Q}{x-z_1} + \left(\frac{1}{x-z_1}-\frac{1}{x-z_2}\right) H(u) \ , \ & u&=\frac{(x-z_1)z_{23}}{(x-z_3)z_{21}}\ ,
\label{wox}
\\
W_{m\geq 2}(x_1,\cdots x_m) &= \prod_{j=1}^m \left(\frac{1}{x_j-z_1}-\frac{1}{x_j-z_2}\right) H(u_1,\cdots u_m) \ , \  & u_j&=\frac{(x_j-z_1)z_{23}}{(x_j-z_3)z_{21}}\ ,
\label{wmx}
\end{align}
where $H(u)$ and $H(u_1,\cdots u_m)$ are arbitrary functions. These expressions are symmetric under permutations of $(z_1,z_2,z_3)$, although this is not manifest. 

In the case $n=3$, equations (\ref{wox})-(\ref{wmx}) determine the dependence of $W_m$ not only on $(x_1,\cdots x_m)$, but also on $(z_1,z_2,z_3)$, since there are no coordinates $z_i$ apart from $(z_1,z_2,z_3)$ for forming conformally-invariant cross-ratios $\frac{(z_i-z_1)z_{23}}{(z_i-z_3)z_{21}}$.
This allows us to express the action  of the differential operator $D_x$ (\ref{dx}) on $W_m$ as 
\begin{align}
 D_x W_m(x_1,\cdots x_m)\ \underset{(n=3)}{=}\ -\delta_{m,1}\frac{Q}{\varphi(x)}+\sum_{j=1}^m \pp{x_j}\left(\frac{\varphi(x_j)-\varphi(x)}{(x-x_j)\varphi(x)} W_m(x_1,\cdots x_m) \right)\ .
\label{dxw}
\end{align}
This can be used for eliminating the $D_xW_m(I)$ term in the Ward identity (\ref{qpw}), with the result 
\begin{multline}
 Q\pp{x} W_{m+1}(I,x) -\sum_{J\subset I}W_{|J|+1}(J,x)W_{m-|J|+1}(I-J,x) - W_{m+2}(I,x,x)
\\
 \underset{(n=3)}{=}\ \delta_{m,0}t(x)-\delta_{m,1}\frac{Q}{\varphi(x)} +\sum_{j=1}^m\pp{x_j}\frac{\varphi(x_j)W_m(I)-\varphi(x)W_m(I-\{x_j\},x)}{(x-x_j)\varphi(x)}\ ,
\label{qpwd}
\end{multline}

\subsection{Expansion of the Ward identities \label{subsecewi}}

To directly solve the Ward identities would be difficult. Instead, we will use the perturbative expansion (\ref{ci}) around the heavy asymptotic limit. Actually, there are several possible expansions around this limit. Our choice of $Q$ as the expansion parameter, and of $\eta_i=\frac{\al_i}{Q}$ as the fixed quantities, will lead to simple expansions for the Ward identities, but not necessarily for their solutions. Another possible choice would be to take $b$ such that $Q=b+b^{-1}$ as the small expansion parameter, and to have $b\al_i$ fixed. These two choices are related by a perturbative resummation of the expansion. 

We assume that the expansions of $F=\log Z$, $W_m$ and $t(x)$ take the following forms:
\begin{align}
 F &= \sum_{g=0}^\infty Q^{2-2g} F^{(g)} \ ,
\label{fs}
\\
W_m & = \sum_{g=0}^\infty Q^{2-m-2g} W_m^{(g)}\ ,
\label{ws}
\\
t(x) &= \sum_{g=0}^\infty Q^{2-2g} t^{(g)}(x)\ .
\label{ts}
\end{align}
The behaviour of $W_m$ corresponds to the assumption that the spin one chiral field $J$ is of order $Q$, which is consistent with its behaviour at infinity eq. (\ref{tiqi}) . This implies that $\hat{W}_m$ (\ref{w}) is of order $Q^m$. The behaviour of the connected correlation function $W_m$ is then suggested by its definition (\ref{wii}) from $\hat{W}_m$. The behaviour of $t(x)$ (\ref{tx}) is suggested by the behaviour of the 
conformal dimension $\Delta_{\al}$ (\ref{da}),
\begin{align}
 \Delta_{\al} = Q^2 \delta \scs \delta = \eta(1-\eta)\ .
\label{dad}
\end{align}
At a given order $g\geq 0$, the Seiberg-Witten equation (\ref{swp}) reduces to 
\begin{align}
 \left(\pp{\eta_j}-\pp{\eta_i}\right) F^{(g)}= \int_{z_i}^{z_j} \left(W_1^{(g)}(\overset{i}{x})+W_1^{(g)}(\overset{j}{x})\right)dx\ . 
\label{swg}
\end{align}
It is straighforward to expand the Ward identity (\ref{qpw}) in powers of $Q$, and the $O(Q^{2-2g-m})$ term of that identity is the perturbative Ward identity
 \begin{multline}
 \pp{x} W_{m+1}^{(g)}(I,x) -\sum_{J\subset I}\sum_{h=0}^g W^{(h)}_{|J|+1}(J,x)W^{(g-h)}_{m-|J|+1}(I-J,x) - W^{(g-1)}_{m+2}(I,x,x)
\\
= \delta_{m,0}t^{(g)}(x) + D_xW^{(g)}_m(I)+\sum_{j=1}^m\pp{x_j}\frac{W^{(g)}_m(I)-W^{(g)}_m(I-\{x_j\},x)}{x-x_j}\ .
\label{pwh}
\end{multline}
This identity can be used for determining $W_{m+1}^{(g)}$, provided we already know $W_{m'+1}^{(g')}$ for any pair $(g',m')$ such that 
\begin{align}
 \left(g'< g \quad {\rm and}\quad m'\leq m+g-g'\right) \quad {\rm or} \quad \left( g'=g \quad {\rm and }\quad  m'<m\right)\ .
\end{align}
Therefore, the Ward identities allow us to recursively determine all $W_m^{(g)}$. 
This is what we now illustrate at low orders. 
The first two orders ($g=0,1$) of the expansion of the first Ward identity ($m=0$) are 
\begin{align}
\pp{x} W_1^{(0)}(x) - \left(W_1^{(0)}(x)\right)^2 & = t^{(0)}(x) \ ,
\label{wo}
\\
\left(\pp{x}-2W_1^{(0)}(x)\right)W_1^{(1)}(x) -W_2^{(0)}(x,x) & = t^{(1)}(x) \ .
\label{wi}
\end{align}
The leading order ($g=0$) of the second Ward identity ($m=1$) reads
\begin{align}
 \left(\pp{x}-2W_1^{(0)}(x)\right) W_2^{(0)}(x,y) = D_x W^{(0)}_1(y) + \pp{y}\frac{W^{(0)}_1(y)-W^{(0)}_1(x)}{x-y}\ .
\label{wt}
\end{align}
Considering $t^{(0)}(x)$ and $t^{(1)}(x)$ as known, we can solve eq. (\ref{wo}) for $W_1^{(0)}$, then eq. (\ref{wt}) for $W_2^{(0)}$, and then eq. (\ref{wi}) for $W_1^{(1)}$. We could in principle go on and use equations with higher $m$ and $g$, so as to determine successively $W_3^{(0)}$, $W_2^{(1)}$, $W_1^{(2)}$, etc. 

Let us comment on the equation (\ref{wo}) for $W_1^{(0)}$. Introducing the function $\Psi$ such that
\begin{align}
 W_1^{(0)} = -\frac{\Psi'}{\Psi}\ ,
\label{wpp}
\end{align}
the equation for $W_1^{(0)}$ becomes a second-order linear differential equation for $\Psi$,
\begin{align}
 \left(\p^2 + t^{(0)}\right) \Psi = 0 \ .
\label{ptp}
\end{align}
This differential equation illustrates the multivalued nature of the spin one chiral field $J$. In particular, in the neighbourhood of $x=z_i$, we have $t^{(0)}(x) = \frac{\eta_i(1-\eta_i)}{(x-z_i)^2} + O(\frac{1}{x-z_i})$, which leads to two solutions $\Psi(\overset{i+}{x})$ and $\Psi(\overset{i-}{x})$ corresponding to the two critical exponents $\eta_i$ and $1-\eta_i$,
\begin{align}
 \Psi(\overset{i+}{x}) = (x-z_i)^{\eta_i}(1+O(x-z_i)) \scs  \Psi(\overset{i-}{x}) = (x-z_i)^{1-\eta_i}(1+O(x-z_i))\ .
\label{pip}
\end{align}
The corresponding solutions for $W_1^{(0)}$ are the only solutions of eq. (\ref{wo}) which are meromorphic in the neighbourhood of $x=z_i$, and they behave as 
\begin{align}
 W_1^{(0)}(\overset{i+}{x}) = \frac{-\eta_i}{x-z_i} + O(1) \scs W_1^{(0)}(\overset{i-}{x}) = \frac{\eta_i-1}{x-z_i} + O(1) \ .
\label{wpm}
\end{align}
This closely parallels the "multivalued operator product expansion" $J(x) V_\al(z,\bar{z})$ eq. (\ref{jvjv}). 

A generic solution $W_1^{(0)}$ of eq. (\ref{wo}), built from an arbitrary combination $\Psi = \sum_\pm \lambda_\pm \Psi(\overset{i\pm}{x})$, 
must have a branch cut ending at $z_i$. We will use the requirement of meromorphy near $x=z_i$ for defining particular solutions of Ward identities, thereby solving the problem of the initial condition in these first-order differential equations. Of course, the resulting solutions still depend on the choices of the point $z_i$ and of the critical critical exponent $\eta_i$ or $1-\eta_i$. These choices will be indicated by superscripts  as in $W_1^{(0)}(\overset{i+}{x})$.

\section{Perturbative calculation of the three-point function \label{sectpf}}

We now want to solve the Ward identities, in order to derive the Liouville three-point function, which in our notations is the $n=3$ case of the partition function $Z$ (\ref{z}). We will focus on this case because the Liouville three-point function is explicitly known, and given by a non-trivial yet reasonably simple formula. (See Section \ref{subsecltp}.) 

In the $n=3$ case, some simplifications occur in our Ward identities. First of all, we can use the form (\ref{qpwd}) of the Ward identities, where the $D_xW_m(I)$ term has been eliminated thanks to global conformal symmetry. Moreover, the form (\ref{tf}) for $t(x)$, together with the behaviour (\ref{dad}) of the conformal dimensions $\Delta_{\al_i}$, imply that $t(x)$ 
vanishes beyond the leading order:
\begin{align}
 t(x) \ \underset{(n=3)}{=}\ Q^2 t^{(0)}(x)\ .
\label{tz}
\end{align}
In the rest of this Section, we will omit the indication "$(n=3)$".

We will compute the first two terms $F^{(0)}$ and $F^{(1)}$ of the perturbative expansion (\ref{fs}) of $F=\log Z$. To compute higher order terms looks complicated although in principle doable.

\subsection{Calculation of $F^{(0)}$}

The task of computing $F^{(0)}$ now boils down to solving the second-order equation (\ref{ptp}), deducing $W_1^{(0)}$ thanks to eq. (\ref{wpp}), and using the Seiberg-Witten equation (\ref{swg}) for obtaining $F^{(0)}$. 

The second-order equation (\ref{ptp}) can be solved explicitly in the case $n=3$, because it amounts to a hypergeometric equation. For convenience, we reproduce the equation:
\begin{align}
\left(\p^2 + t^{(0)}\right) \Psi = 0 \scs  t^{(0)}(x) = \frac{1}{\prod_{i=1}^3 (x-z_i)} \sum_{i=1}^3 \delta_i \frac{z_{ij}z_{ik}}{x-z_i}\ ,
\label{tzx}
\end{align}
where we assume $i\neq j\neq k$, and we recall the definition $\delta_i=\eta_i(1-\eta_i)$. 
We have introduced a labelling (\ref{pip}) of particular solutions, based on their behaviour near the critical points $z_i$. 
We will now assume that the parameters obey 
\begin{align}
 \Re \eta_i < \frac12\ ,
\label{re}
\end{align}
so that the solution $\Psi(\overset{i+}{x})$ is dominant with respect to the solution $\Psi(\overset{i-}{x})$. We will work with dominant solutions only, and simplify the notations by writing $\Psi(\overset{i+}{x})=\Psi(\overset{i}{x})$.
For example, the dominant solution near $x=z_1$ is
\begin{align}
 \Psi(\overset{1}{x}) =(x-z_1)^{\eta_1} \left(\frac{x-z_2}{z_{12}}\right)^{\eta_2}\left(\frac{x-z_3}{z_{13}}\right)^{1-\eta_1-\eta_2} F\left(\eta_{12}^3,\eta_{123}-1;2\eta_1;\frac{(x-z_1)z_{23}}{(x-z_3)z_{21}}\right)\ ,
\label{pox}
\end{align}
where we use the notations
\begin{align}
 \eta_{12}^3 = \eta_1+\eta_2-\eta_3 \scs \eta_{123} = \eta_1+\eta_2+\eta_3\ .
\end{align}
The behaviour of $\Psi(\overset{i}{x})$ near $x=z_j$ with $j\neq i$ is 
\begin{align}
 \Psi(\overset{i}{x}) = c_{ij} (x-z_j)^{\eta_j}\left(1+O((x-z_j)^{1-2\eta_j}+O(x-z_j))\right)\ ,
\label{pxc}
\end{align}
where the coefficient $c_{ij}$ is an element of the monodromy matrix of the hypergeometric equation, which can be deduced from eq. (\ref{fff}),
\begin{align}
 c_{12} = \frac{\G(2\eta_1)\G(1-2\eta_2)}{\G(\eta_{13}^2)\G(1-\eta_{23}^1)} z_{12}^{\eta_1-\eta_2}\left(\frac{z_{13}}{z_{23}}\right)^{\eta_1+\eta_2-1}\ .
\label{cot}
\end{align}
This is all the information we will need on the solution $\Psi(\overset{i}{x})$. The corresponding solution for $W_1^{(0)}(x)$ will be denoted 
$W_1^{(0)}(\overset{i}{x}) = -\pp{x} \log \Psi(\overset{i}{x})$. It is meromorphic in the neighbourhood of $z_i$ and has a branch cut on $(z_j,z_k)$ where $i\neq j\neq k$.

Using the perturbative Seiberg-Witten equation (\ref{swg}) at the order $g=0$, we compute
\begin{align}
 \left(\pp{\eta_j}-\pp{\eta_i}\right) F^{(0)} =  \log \frac{\Psi(\overset{i}{z_i})\Psi(\overset{j}{z_i})}{\Psi(\overset{i}{z_j})\Psi(\overset{j}{z_j})} =  \log\frac{c_{ji}}{c_{ij}}\ ,
\end{align}
where we use the regularization which consists in ignoring the divergent factor $\left. (x-z_i)^{\eta_i}\right|_{x=z_i}$ in $\Psi(\overset{i}{z_i})$. (This regularization is justified in \cite{zz95}.) Using the formula (\ref{cot}) for $c_{ij}$, we obtain
\begin{align}
 \left(\pp{\eta_2}-\pp{\eta_1}\right) F^{(0)} = 2(\eta_2-\eta_1)\log z_{12} +2(\eta_1+\eta_2-1) \log\frac{z_{23}}{z_{13}}+\log 
\frac{\gamma(2\eta_2)\gamma(\eta_{13}^2)}{\gamma(2\eta_1)\gamma(\eta_{23}^1)}\ ,
\end{align}
where the special function $\gamma(x)$ is defined in Appendix \ref{appsf}. This suggests that the free energy $F^{(0)}$ which we obtain is related to the known result of Dorn, Otto, Zamolodchikov and Zamolodchikov $F^{(0)}_{DOZZ}$ eq. (\ref{fz}) by
\begin{align}
 F^{(0)}_{DOZZ} = F^{(0)} + \bar{F}^{(0)}\ ,
\label{ffbf}
\end{align}
where the bar indicates the replacement $z_i\rar \bar{z}_i$. That our $F^{(0)}$ depends only on $z_i$ and not $\bar{z}_i$
is an inevitable consequence of using locally holomorphic quantities $W_m$, which is itself necessary for the integral in the Seiberg-Witten equation (\ref{swp}) to be independent from the precise path of integration. 
It is actually possible to derive the full dependence of $F^{(0)}_{DOZZ}$ on $z_i$ and $\bar{z}_i$, by considering single-valued solutions of eq. (\ref{ptp}) instead of our locally holomorphic solutions $\Psi(\overset{i}{x})$ \cite{zz95}. However, that approach has been used only in the heavy asymptotic limit, which is the leading order of our expansion, and it seems difficult to generalize it to higher orders.

\subsection{Calculation of $W_2^{(0)}$  and the Gelfand-Dikii operator}

We will now solve equation (\ref{wt}) for $W_2^{(0)}$. This depends on the choice of a solution for $W_1^{(0)}$. We will take an arbitrary solution $W_1^{(0)}=-\frac{\Psi'}{\Psi}$, built from 
an arbitrary solution $\Psi(x)$ of the second-order equation (\ref{tzx}). Let us introduce differential operators
\begin{align}
 \nabla_x = \p_x +2\frac{\Psi'(x)}{\Psi(x)} \scs \nabla_x^\dagger = \p_x -2\frac{\Psi'(x)}{\Psi(x)} \ .
\label{nn}
\end{align}
Using these operators, and the simplification (\ref{dxw}) which occur in the case $n=3$, the Ward identity (\ref{wt}) can be rewritten as
\begin{align}
 \nabla_x\left(\frac{1}{(x-y)^2}-2W_2^{(0)}(x,y)\right) = -\p_y\nabla_y^\dagger\frac{\varphi(y)}{(x-y)\varphi(x)}\ .
\label{nfw}
\end{align}
One may wonder whether $\nabla_x\nabla_y W_2^{(0)}(x,y)$ is really symmetric under $x\lrar y$ if we compute it using this equation. 
Checking this involves the third-order differential operator
\begin{align}
 G_x = \nabla_x\p_x\nabla_x^\dagger = \p_x^3 +4t^{(0)}(x) \p_x + 2 (t^{(0)})'(x)\ .
\end{align}
This is called the Gelfand-Dikii operator (see for instance \cite{gd78} in \cite{bs95}), and it has the property
\begin{align}
 G_x \frac{\varphi(x)}{(y-x)\varphi(y)} = G_y\frac{\varphi(y)}{(x-y)\varphi(x)}\ ,
\end{align}
which follows from the fact that $\varphi^2 t^{(0)}$ is a polynomial of degree two. This property directly leads to the expected symmetry of $\nabla_x\nabla_y W_2^{(0)}(x,y)$. Notice that the Gelfand-Dikii operator appears in the classical higher equation of motion of Liouville theory \cite{zam03}, which in our notations reads $G_x\Psi(x)^2 =0$. More generally, the classical higher equations of motion of order $m+1$ reads $\left(\prod_{j=-\frac{m}{2}}^{\frac{m}{2}}(\p-2j\frac{\Psi'}{\Psi})\right) \Psi^m = 0$, and we expect that the corresponding differential operator appears in the calculation of $W_m^{(0)}$.

The solutions of eq. (\ref{nfw}) are 
\begin{align}
 W_2^{(0)}(x,y) = \frac{1}{2(x-y)^2} +\frac{1}{2\Psi^2(x)}\p_y\nabla_y^\dagger \int_a^x \frac{d\xi}{\xi-y}\Psi^2(\xi)\frac{\varphi(y)}{\varphi(\xi)}\ ,
\label{wtz}
\end{align}
where $a$ is an a priori $y$-dependent integration bound. This bound is constrained by requiring $W_2^{(0)}$ to be a solution of eq. (\ref{nfw}) with $x\lrar y$. This can be shown to hold if $a=z_i$ for some $i$, provided $\Psi(z_i)=0$, which itself holds if $\Re \eta_i>0$. 

It remains to compute $W_2^{(0)}(x,x)$, which appears in the equation (\ref{wi}) for $W_1^{(1)}$. This is not completely trivial, because 
the regularity of $W_2^{(0)}(x,y)$ at $x=y$ is not manifest in
our formula (\ref{wtz}). We obtain
\begin{align}
 W_2^{(0)}(x,x) = \frac12 \nabla_x\left[\p_x\int_a^x\frac{d\xi}{\xi-x}\left(\frac{\Psi^2(\xi)}{\Psi^2(x)}\frac{\varphi(x)}{\varphi(\xi)}-1\right)-\frac{1}{x-a}\right]+\nabla_x\p_x\log\Psi(x) +2\hat{t}^{(0)}(x)\ ,
\label{wxx}
\end{align}
where we define $\hat{t}^{(0)}(x)$ from $t^{(0)}(x)$ in eq. (\ref{tzx}) by
\begin{align}
 \hat{t}^{(0)}(x) = \left. t^{(0)}(x)\right|_{\delta_i \rar \delta_i-\frac14}\ .
\end{align}

To conclude this section on $W_2^{(0)}$, notice that $W_2^{(0)}(x,y)-\frac{\frac12}{(x-y)^2}$ coincides with the Bergman kernel of \cite{cem09} (Section 4.3 therein), where many of our equations can be found in a more general form.

\subsection{Calculation of $F^{(1)}$}

Let us proceed to the calculation of the solution $W_1^{(1)}(\overset{i}{x})$ of eq. (\ref{wo}) which is holomorphic near $x=z_i$, just like $W_1^{(0)}(\overset{i}{x})$ is meromorphic near $x=z_i$. This solution is formally written as $W_1^{(1)}(\overset{i}{x}) = \nabla_i^{-1} W_2^{(0)}(\overset{i}{x},\overset{i}{x})$, where $W_2^{(0)}(\overset{i}{x},\overset{i}{x})$ is obtained by using the solution $\Psi(\overset{i}{x})$ and the integration bound $a=z_i$ in eq. (\ref{wxx}), and the operator $\nabla_i$ is obtained from $\nabla_x$ (\ref{nn}) by using that same solution $\Psi(\overset{i}{x})$. We thus find
\begin{align}
 W_1^{(1)}(\overset{i}{x}) = \frac12\p_x I(\overset{i}{x}) -\frac{1}{2(x-z_i)} + \p_x\log\Psi(\overset{i}{x}) +2\nabla_i^{-1} \hat{t}^{(0)}(x)\ ,
\label{woo}
\end{align}
where we define
\begin{align}
I(\overset{i}{x})= \int_{z_i}^x\frac{d\xi}{\xi-x}\left(\frac{\Psi^2(\overset{i}{\xi})}{\Psi^2(\overset{i}{x})}\frac{\varphi(x)}{\varphi(\xi)}-1\right)\ .
\label{ix}
\end{align}
Each one of the last three terms of $W_1^{(1)}(\overset{i}{x})$ is meromorphic near $x=z_i$ with a first-order pole at $x=z_i$ (this requirement fixes the ambiguity in $\nabla_i^{-1} \hat{t}^{(0)}(x)$), and the sum of the residues vanishes. The first term $\frac12\p_x I(\overset{i}{x})$ is holomorphic near $x=z_i$. 

We now use the perturbative Seiberg-Witten equation (\ref{swg}) at the order $g=1$, and compute $F^{(1)}$. This involves integrating $W_1^{(1)}(\overset{i}{x})$. The only term in $W_1^{(1)}(\overset{i}{x})$ which does not have an obvious primitive is the last term, namely $\nabla_i^{-1} \hat{t}^{(0)}(x)$. However, it turns out that it is easy to write a primitive for the sum $\nabla_i^{-1} \hat{t}^{(0)}(x)+\nabla_j^{-1} \hat{t}^{(0)}(x)$ which appears in $W_1^{(1)}(\overset{i}{x})+W_1^{(1)}(\overset{j}{x})$,
\begin{align}
 \nabla_i^{-1} \hat{t}^{(0)}(x)+\nabla_j^{-1} \hat{t}^{(0)}(x) = \pp{x}\left(\frac{\Psi(\overset{i}{x})\Psi(\overset{j}{x})}{w_{ij}}\left(\nabla_i^{-1} \hat{t}^{(0)}(x)-\nabla_j^{-1} \hat{t}^{(0)}(x)\right)\right)\ ,
\end{align}
where the constant $w_{ij}$ is defined as the Wronskian
\begin{align}
 w_{ij} = \Psi(\overset{i}{x})\Psi'(\overset{j}{x}) - \Psi'(\overset{i}{x})\Psi(\overset{j}{x}) \ .
\label{wij}
\end{align}
This leads to 
\begin{align}
 \left(\pp{\eta_j}-\pp{\eta_i}\right)F^{(1)} = \left[\frac12 I(\overset{i}{x}) +\frac12\log\frac{\Psi(\overset{i}{x})^2 }{x-z_i} + \frac{2\Psi(\overset{i}{x})\Psi(\overset{j}{x})}{w_{ij}}\nabla_i^{-1}\hat{t}^{(0)}(x)\right]_{z_i}^{z_j} - \left(i\lrar j\right)\ .
\label{ppf}
\end{align}
We will compute this expression by evaluating each term. While the total expression is finite, individual terms can be divergent. While the total expression cannot depend on $(z_1,z_2,z_3)$ due to global conformal symmetry, individual terms can. 

To compute $I(\overset{i}{z_k})$ with $k=i$ or $k=j$, notice that the factor $\varphi(x)$ in the integral expression (\ref{ix}) for $I(\overset{i}{x})$ obeys $\varphi(z_k)=0$. This allows us to replace $\varphi(\xi)$ and $\Psi(\overset{i}{\xi})$ factors with their leading behaviours near $\xi=z_k$. Using the integral formula (\ref{ipip}) we obtain
\begin{align}
 I(\overset{i}{z_i}) = \psi(2\eta_i)-\psi(1) \scs I(\overset{i}{x}) \underset{x\rar z_j}{=} \log\frac{z_{ij}}{x-z_j} + \psi(1-2\eta_j)-\psi(1) +O(x-z_j)\ ,
\end{align}
where the special function $\psi$ is defined in eq. (\ref{pdg}).
The contribution of the term $\frac12\log\frac{\Psi(\overset{i}{x})^2 }{x-z_i}$ in eq. (\ref{ppf}) is easily computed using the behaviours (\ref{pip}) and (\ref{pxc}) of $\Psi(\overset{i}{x})$ near $x=z_i$ and $x=z_j$ respectively. Let us now study the last term in eq. (\ref{ppf}). The integration constant in $\nabla_i^{-1}\hat{t}^{(0)}(x)$ is determined by the requirement that this function be meromorphic near $x=z_i$, and this leads to
\begin{align}
 \nabla_i^{-1}\hat{t}^{(0)}(x) = \frac{1}{\Psi(\overset{i}{x})^2}\int_{z_i}^x d\xi\ \Psi(\overset{i}{\xi})^2 \hat{t}^{(0)}(\xi)\ ,
\end{align}
where the power-like divergence at $\xi=z_i$ is regularized by analytic continuation in the parameter $\eta_i$. 
We thus have $\left.\frac{\Psi(\overset{i}{x})\Psi(\overset{j}{x})}{w_{ij}}\nabla_i^{-1}\hat{t}^{(0)}(x)\right|_{x=z_i} = 0$. On the other hand,
$\nabla_i^{-1}\hat{t}^{(0)}(x)$ has a logarithmic divergence at $x=z_j$, which cancels the logarithmic divergences of the other terms of eq. (\ref{ppf}), and which  we will regularize by introducing a factor of $(\xi-z_j)^{\epsilon_j}$ in the integrand. Keeping this regularization implicit, we have 
\begin{align}
 \left.\frac{\Psi(\overset{i}{x})\Psi(\overset{j}{x})}{w_{ij}}\nabla_i^{-1}\hat{t}^{(0)}(x)\right|_{x=z_j} = \frac{\int_{z_i}^{z_j} d\xi\ \Psi(\overset{i}{\xi})^2 \hat{t}^{(0)}(\xi)}{c_{ij}w_{ij}} \overset{\begin{array}{c} \rm definition \end{array}}{\underset{ \begin{array}{c} \scriptstyle (i,j,k)=(1,2,3)\\ \vspace{-7mm} \\ \scriptstyle (z_1,z_2,z_3)=(0,1,\infty) \end{array}}{=}} E(\eta_{12}^3,\eta_{123}-1,2\eta_1)\ .
\label{ppd}
\end{align}
So, by sending $(z_1,z_2,z_3)$ to $(0,1,\infty)$, we defined a function $E(a,b,c)$ which depends only on $(\eta_1,\eta_2,\eta_3)$, although it is more convenient to take $(\eta_{12}^3,\eta_{123}-1,2\eta_1)$ as its arguments. 

We know the coefficient $c_{ij}$ (\ref{cot}), and we can compute the Wronskian $w_{ij}$
(\ref{wij})
with the help of the transformation property (\ref{fff}) of the hypergeometric function, 
\begin{align}
 w_{12} = \frac{\G(2\eta_1)\G(2\eta_2)}{\G(\eta_{12}^3)\G(\eta_{123}-1)} z_{12}^{\eta_1+\eta_2-1} \left(\frac{z_{13}}{z_{23}}\right)^{\eta_1-\eta_2}\ .
\end{align}
This leads to the integral representation (\ref{ei}) for the function $E(a,b,c)$. 
Appendix \ref{appih} is devoted to computing that integral. Combining the result with the other terms of eq. (\ref{ppf}), we obtain 
\begin{multline}
  \left(\pp{\eta_2}-\pp{\eta_1}\right)F^{(1)} =\log 
\frac{\gamma(2\eta_1)\gamma(\eta_{23}^1)}{\gamma(2\eta_2)\gamma(\eta_{13}^2)}
\\
+ (\eta_{13}^2-\tfrac12)\left(\psi(\eta_{13}^2)+\psi(1-\eta_{13}^2)\right) -(\eta_{23}^1-\tfrac12)\left(\psi(\eta_{23}^1+\psi(1-\eta_{23}^1)\right) 
\\
+(2\eta_2-\tfrac12)\left(\psi(2\eta_2)+\psi(1-2\eta_2)\right) -(2\eta_1-\tfrac12)\left(\psi(2\eta_1)+\psi(1-2\eta_1)\right)
\ .
\end{multline}
This suggests that our $F^{(1)}$ is related to the known result $F^{(1)}_{DOZZ}$ eq. (\ref{fg}) by 
\begin{align}
 F^{(1)}_{DOZZ} = F^{(1)}+\bar{F}^{(1)}= 2 F^{(1)}\ .
\label{fo}
\end{align}

\section{Conclusions}

\subsection{The results and their possible generalizations}

We have shown that our results for the first two terms $F^{(0)}$ and $F^{(1)}$ of the (logarithm of the) Liouville three-point function  are consistent with already known results. One may object that our Seiberg-Witten equation (\ref{swg}) did not allow us to compute $F^{(g)}$, but rather differences of derivatives $\left(\pp{\eta_i}-\pp{\eta_j}\right)F^{(g)}$, thereby leaving a large indeterminacy. However, this indeterminacy is lifted if we assume that the reflection symmetry (\ref{vrv}) holds, even if we do not know the reflection coefficient, as explained in Appendix \ref{apprs}. Therefore, we can compute $F^{(0)}$ and $F^{(1)}$ up to unimportant normalization constants. 

We expect that all $F^{(g\geq 2)}$ can similarly be computed, and agree with the known results from Dorn, Otto, Zamolodchikov and Zamolodchikov as in the cases of $F^{(0)}$ and $F^{(1)}$, eq. (\ref{ffbf}) and eq. (\ref{fo}). Thus, we expect the non-perturbative relations
\begin{align}
 F_{DOZZ} = F+\bar{F} \quad {\rm or\ equivalently} \quad Z_{DOZZ} = |Z|^2\ .
\label{zzz}
\end{align}
This is analogous to the holomorphic factorization of black hole partition functions into topological string partition functions found by Ooguri, Strominger and Vafa \cite{osv04}, in the sense that our calculation of $Z$ involves integrals of locally holomorphic quantities -- that is, topological objects in two dimensions. 

Completely solving Liouville theory would mean computing not only the three-point function, but also general $n$-point functions, which do not holomorphically factorize. Such $n$-point functions can however be reduced to  
locally holomorphic quantities, by using their decomposition into combinations of three-point functions and conformal blocks. 
In order to compute an $n$-point conformal block, we would need to deal with the existence of $n$ quantum accessory parameters (\ref{be}), of which $n-3$ are not fixed by conformal symmetry. However, an $n$-point conformal block depends not only on the parameters of the $n$ involved fields $V_{\al_i}(z_i)$, but also on
$n-3$ extra parameters, namely the conformal dimensions in the internal channels. These extra parameters are expected to be related to the quantum accessory parameters. 

We expect our methods to work not only in Liouville theory, but also in other conformal field theories. We certainly need these theories to be non-rational, so that we can write Seiberg-Witten equations like eq. (\ref{sw}) using continuous parameters like $\al_i$.
It seems possible to generalize our construction to conformal Toda theory, although this will not necessarily lead to the explicit computation of the three-point function, which is a difficult problem already at the leading order \cite{fl07c}. It would be interesting to generalize our construction to Wess-Zumino-Novikov-Witten models. In such models there are meromorphic spin one currents $J^a(z)$, and one may be tempted to use them in Seiberg-Witten equations as analogs of our chiral field $J(z)$. But a preliminary analysis shows that we actually need multivalued spin one chiral fields, which do not coincide with $J^a(z)$. 

\subsection{The non-commutative spectral curve \label{subsecsc}}

Here we discuss the interpretation of our calculations in terms of the spectral curve whose equation is eq. (\ref{yt}). This interpretation is unfortunately not precise enough for justifying the form of the Seiberg-Witten equation (\ref{swp}). A comprehensive geometrical framework for non-commutative spectral curves has been developed in \cite{cem09}, but it makes important technical assumptions which our curves do not obey.

The spectral curve is supposed to be determined not only by the theory under consideration, but also by the particular expansion we want to use for the correlation functions. The leading order of the expansion is supposed to provide the equation of the curve, and the higher orders are supposed to be computed from certain geometrical quantities of that same curve. In principle, we may refrain from doing any expansion and consider the associated "non-perturbative spectral curve", but determining this curve would be equivalent to computing the partition function, and the curve would therefore be useless as a computational tool. So we want to consider an expansion of the partition function in terms of one or several parameters. The more parameters, the simpler the curve, and the more complicated the expansion. 

In the case of Liouville theory, there is only one natural expansion parameter -- the central charge, and we have demonstrated that the corresponding curve and the associated quantities $W_m^{(g)}$ are tractable. We could nevertheless perform a further expansion with respect to a second parameter. In particular we could introduce a parameter $h$ such that we would have $[y,x]=h$ instead of $[y,x]=1$ in the equation (\ref{yt}) of the spectral curve. The resulting spectral curve, which would by definition be obtained in the limit $h\rar 0$, would then be commutative. But the expansion would then be a double series in $Q^2$ and $h$, which is an unnecessary complication as far as the calculation of the free energy is concerned. 

So, we consider the non-commutative spectral curve which is defined by the second-order differential equation (\ref{yt}). By analogy with the corresponding commutative spectral curve, the existence of two linearly independent solutions of that equation can be interpreted as the existence of two sheets above a point $x$. Since the solutions can be linearly combined, the definition of such sheets is rather arbitrary. However, we have found that near each singularity $x=z_i$ of the curve, it is possible to define two priviledged sheets corresponding to the two solutions with diagonal monodromy transformations, so that the corresponding correlation functions $W_1^{(0)}(\overset{i\pm}{x})$ are meromorphic near $x=z_i$. Such priviledged solutions play a crucial role in lifting ambiguities in Seiberg-Witten equations and in solutions of the Ward identities. But their existence relies on the particular form of the spectral curve, which is why our calculations are difficult to interpret in terms of a general theory 
of non-commutative spectral curves.

Such an interpretation would nevertheless be very interesting, because we might then access such a powerful result as the invariance of the free energy $F$ under symplectic transformations of the spectral curve, which is known to hold in the case of commutative spectral curves \cite{eo08}. This could then lead to simple proofs of relations between apparently different theories, such as the AGT relation.

\appendix

\section{Useful technical results}

\subsection{Special functions \label{appsf}}

We assume that Euler's Gamma function $\Gamma(x)$ such that $\G(x+1)=x\G(x)$ is known. This function appears in the integral 
\begin{align}
 \int_0^1 x^\al (1-x)^\beta dx= \frac{\G(\al+1)\G(\beta+1)}{\G(\al+\beta+2)}\ .
\label{izo}
\end{align}
The function $\gamma(x)$ is defined as a ratio of two Gamma functions:
\begin{align}
 \gamma(x) = \frac{\G(x)}{\G(1-x)}\ .
\end{align}
The function $\psi(x)$ is defined as
\begin{align}
 \psi(x) = \pp{x} \log \Gamma(x)\ .
\label{pdg}
\end{align}
This implies in particular 
\begin{align}
 \pp{x} \log \gamma(x) = \psi(x)+\psi(1-x)\ .
\end{align}
The function $\psi(x)$ has the series representation
\begin{align}
 \sum_{n=0}^\infty \frac{1}{(x+n)(y+n)} = \frac{\psi(x)-\psi(y)}{x-y}\ .
\label{sxy}
\end{align}
The function $\psi(x)$ also has integral representations,
\begin{align}
  \int_0^1 du \frac{u^\lambda-1}{u-1} =\psi(\lambda+1)-\psi(1) \scs \int_1^\frac{1}{\e} du\frac{u^\lambda-1}{u-1} =\log\e+\psi(1) -\psi(-\lambda) + o(1)\ .
\label{ipip}
\end{align}
We review some well-known properties of the hypergeometric function. First of all, we have the series representation
\begin{align}
 F(a,b;c;x) = \frac{\G(c)}{\G(a)\G(b)}\sum_{n=0}^\infty \frac{\G(a+n)\G(b+n)}{n!\G(c+n)} x^n\ .
\label{fsum}
\end{align}
We have the transformation property
\begin{multline}
F(a,b;c;x) = \frac{\G(c)\G(c-a-b)}{\G(c-a)\G(c-b)}F(a,b;a+b-c+1;1-x)
\\
 + (1-x)^{c-a-b} \frac{\G(c)\G(a+b-c)}{\G(a)\G(b)}F(c-a,c-b;c-a-b+1;1-x)\ .
\label{fff} 
\end{multline}
The last two formulas imply
\begin{align}
 \sum_{m=0}^\infty \frac{\G(a+m)\G(b+m)}{m!\G(c+m)} = \frac{\G(a)\G(b)\G(c-a-b)}{\G(c-a)\G(c-b)}\ .
\label{sgg}
\end{align}
Another useful transformation of the hypergeometric function is 
\begin{align}
 F(a,b;c;x) = (1-x)^{c-a-b} F(c-a,c-b;c;x)\ .
\label{fpf}
\end{align}

\subsection{Expansion of the Liouville three-point function \label{subsecltp}}

We recall the Dorn-Otto-Zamolodchikov-Zamolodchikov formula \cite{do94,zz95} for the three-point function $Z_{DOZZ} = \la V_{\al_1}(z_1)V_{\al_2}(z_2)V_{\al_3}(z_3)\ra$ of Liouville theory at central charge $c$:
\begin{multline}
 Z_{DOZZ}= |z_{12}|^{2(\Delta_{\al_3}-\Delta_{\al_1}-\Delta_{\al_2})} |z_{23}|^{2(\Delta_{\al_1}-\Delta_{\al_2}-\Delta_{\al_3})} |z_{13}|^{2(\Delta_{\al_2}-\Delta_{\al_1}-\Delta_{\al_3})}
\\
\times \lambda\ \mu^{\sum_{i=1}^3 \al_i} \frac{\up(2\al_1)\up(2\al_2)\up(2\al_3)}{\up(\al_{123}-Q)\up(\al_{12}^3)\up(\al_{23}^1)\up(\al_{13}^2)} \ ,
\label{zozz}
\end{multline}
where $c=1+6Q^2$ and $Q=b+b^{-1}$. Here $\lambda$ and $\mu$ are $\al_i$-independent normalization constants, which can be determined neither in our approach (see Appendix \ref{apprs}) nor in the conformal bootstrap approach. The special function $\up(x)$, which depends on the parameter $b$ and is invariant under $b\rar b^{-1}$, is defined by
\begin{align}
 \log \up(x) = \int_0^\infty \frac{dt}{t}\left[\left(\frac{Q}{2}-x\right)^2e^{-t} -\frac{\sinh^2\left(\frac{Q}{2}-x\right)\frac{t}{2}}{\sinh\frac{bt}{2}\sinh\frac{t}{2b}}\right]\ .
\end{align}
Let us compute the expansion of $F_{DOZZ}=\log Z_{DOZZ}$ around the heavy asymptotic limit (\ref{ci}).  To begin with, we compute the expansion of $\log\up(Q\eta)$ with  $\eta$ fixed. In this computation, we neglect the $\eta$-independent terms as well as the terms in $(\eta-\frac12)^2$, because such terms contribute to the normalization constants $\lambda,\mu$ in $Z_{DOZZ}$. 
Up to such terms, we find that $\log\up(Q\eta)$ has an expansion in powers of $Q^{-2}$, where the leading term is of order $Q^2$, as in eq. (\ref{fs}),
\begin{align}
 \log\up(Q\eta) = \sum_{g=0}^\infty Q^{2-2g} u^{(g)}(\eta) \ ,
\end{align}
We find
\begin{align}
 u^{(0)}(\eta) &= \upsilon(\eta)\ , \\
 u^{(1)}(\eta) &= -2\upsilon(\eta)+(\eta-\tfrac12)\upsilon'(\eta)\ , \\
 u^{(2)}(\eta) &= -\upsilon(\eta) -(\tfrac12\eta(1-\eta)-\tfrac{1}{12})\upsilon''(\eta)\ ,
\end{align}
where we use the function
\begin{align}
 \upsilon(\eta) = \int_{\frac12}^\eta \log\gamma(t) dt\ .
\end{align}
In our expansion, the $z_i$-dependent prefactors of $Z_{DOZZ}$ contribute only at the leading order, and we have
\begin{align}
 F^{(0)}_{DOZZ} &= T_{\{\eta_i\}}(u^{(0)})-\delta_{12}^3\log |z_{12}|^2 -\delta_{23}^1\log |z_{23}|^2 - \delta_{13}^2 \log |z_{13}|^2   \ ,
 \label{fz}
\\ 
F^{(g\geq 1)}_{DOZZ} & = T_{\{\eta_i\}} (u^{(g)})\ ,
\label{fg}
\end{align}
where we define the functional
\begin{align}
 T_{\{\eta_i\}}(f) = f(2\eta_1)+f(2\eta_2)+f(2\eta_3)-f(\eta_{12}^3)-f(\eta_{23}^1)-f(\eta_{13}^2) -f(\eta_{123}-1)\ .
\end{align}

\subsection{Calculation of integrals involving hypergeometric functions \label{appih}}

The following integral appears in the calculation of $F^{(1)}$, more precisely in eq. (\ref{ppd}):
\begin{multline}
 E(a,b,c) =  -\frac{\G(a)\G(b)\G(c-a)\G(c-b)}{4\G(c)^2\G(c-a-b)\G(a+b-c+1)} 
\\
\times \int_0^1 d\xi\ \xi^{\e_1+c-1}(1-\xi)^{\e_2+a+b-c} F(a,b;c;\xi)^2 \left[\frac{(c-1)^2}{\xi}+\frac{(a+b-c)^2}{1-\xi}-(a-b)^2\right] \ ,
\label{ei}
\end{multline}
where we introduce two regulators $\e_1$ and $\e_2$. 
Let us decompose it into three terms, as suggested by the last factor:
\begin{align}
 E(a,b,c) = -\frac14 \left[(c-1)E_1 + (a+b-c)E_2 +(a-b)E_3\right]\ .
\end{align}
The regulators $\e_1$ and $\e_2$ are needed for the terms $E_1$ and $E_2$ respectively, and will be set to zero in the other terms.
In each of the three terms $E_1,E_2$ and $E_3$ we expand one factor $F(a,b;c;\xi)$ in powers of $\xi$, and the other factor $F(a,b;c;\xi)$ in powers of $(1-\xi)$, using the formulas (\ref{fsum}) and (\ref{fff}) for the hypergeometric function. Then we perform the integral over $\xi$ using eq. (\ref{izo}). The result is 
\begin{multline}
 E_1 = (c-1)\sum_{m,n=0}^\infty \frac{\G(c-1+m+\e_1)}{m!\G(c+m)}\left(\frac{\G(a+m)\G(b+m)}{\G(a)\G(b)} 
\frac{\G(a+n)\G(b+n)}{n!\G(a+b+m+n+\e_1)} \right.
\\
- \left. \frac{\G(c-a+m)\G(c-b+m)}{\G(c-a)\G(c-b)}\frac{\G(c-a+n)\G(c-b+n)}{n!\G(2c-a-b+m+n+\e_1)}\right) \ ,
\label{eo}
\end{multline}
where in the second term we used the transformation (\ref{fpf}) before expanding the hypergeometric function in powers of $\xi$,
\begin{multline}
 E_2 =\frac{a+b-c}{\G(a)\G(b)}\sum_{m,n=0}^\infty \left(\frac{\G(a+m)\G(b+m)}{m!\G(a+b+m+n+\e_2)} \frac{\G(a+n)\G(b+n)}{n!}\frac{\G(a+b-c+n+\e_2)}{\G(a+b-c+1+n)} \right.
\\  - \left.\frac{\G(a+m)\G(b+m)}{m!\G(c+m+n+\e_2)}
\frac{\G(c-a+n)\G(c-b+n)}{\G(c-a-b+1+n)}\frac{\G(n+\e_2)}{n!}\right)\ ,
\end{multline}
\begin{multline}
 E_3 = \frac{b-a}{\G(a)\G(b)} \sum_{m,n=0}^\infty \frac{\G(a+m)\G(b+m)}{m!}
\\
\times\left(\frac{\G(a+n)\G(b+n)}{n!\G(a+b+1+m+n)} - \frac{\G(c-a+n)\G(c-b+n)}{\G(c+1+m+n)\G(c-a-b+1+n)}\right)\ .
\end{multline}
These sums can be performed, and expressed in terms of the $\psi$ function eq. (\ref{pdg}),
\begin{align}
 E_1 &= \psi(c-a)+\psi(c-b)-\psi(a)-\psi(b)\ ,
\\
E_2 & = \frac{2}{\e_2} +2\psi(a+b-c)+2\psi(c-a-b) -\psi(c-a)-\psi(c-b)-\psi(a)-\psi(b) + O(\e_2)\ ,
\\
E_3 & = \psi(b)+\psi(c-b)-\psi(a)-\psi(c-a)\ ,
\end{align}
so that we have 
\begin{multline}
 2E(a,b,c) = (a-\tfrac12)\psi(a) + (b-\tfrac12)\psi(b) +(a-c+\tfrac12)\psi(c-a)+(b-c+\tfrac12)\psi(b-c) 
\\
+ (c-a-b)\left(\psi(a+b-c)+\psi(c-a-b) +\frac{1}{\epsilon_2}\right) +O(\epsilon_2)\ .
\end{multline}
We now give some details on the computations which lead to these results.

\paragraph{Computation of $E_3$.} 

In each term of $E_3$, we use the formula (\ref{sgg}) for performing the sum over $m$. The sum over $n$ is then of the type (\ref{sxy}), and directly produces the $\psi$ functions.

\paragraph{Computation of $E_2$.}

In each term of $E_2$, we use the formula (\ref{sgg}) for performing the sum over $m$. Then we separately consider the term $n=0$ and the rest of the sum. The term $n=0$ diverges as $\e_2\rar 0$ due to a $\G(\e_2)$ factor, and we have to expand the remaining Gamma functions in powers of $\e_2$ up to $O(\e_2)$, using $\frac{\G(x+\e)}{\G(x)} = 1 + \e\psi(x) + O(\e^2)$. The $n\geq 1$ terms all have a finite limit as $\e_2\rar 0$, and their sum is of the type (\ref{sxy}).

\paragraph{Computation of $E_1$.}

In both terms of $E_1$, we use the formula (\ref{sgg}) for performing the sum over $n$. The rest of the calculation is analogous to the calculation of $E_2$, with the $m=0$ term diverging as $\e_1\rar 0$, and the rest of the sum being of the type (\ref{sxy}). 
Combining the two terms of $E_1$, the divergent $O(\frac{1}{\e_1})$ terms  cancel.

\subsection{Reflection symmetry \label{apprs}}

Let us show that the knowledge of the derivatives 
\begin{align}
 f_{ij}=\left(\pp{\al_i}-\pp{\al_j}\right)F\ ,
\label{fij}
\end{align}
which appear in our Seiberg-Witten equations (\ref{swp}), together with the assumption of reflection symmetry (\ref{vrv}), determine $F$ up to normalization constants. At the level of the free energy, the reflection symmetry implies 
\begin{align}
 F(\al_1,\al_2,\cdots) = F(Q-\al_1,\al_2,\cdots) + r(\al_1) \ ,
\end{align}
for some unknown function $r(\al)$. This implies in particular 
\begin{align}
 f_{12}(Q-\al_1,\al_2,\cdots) +f_{12}(\al_1,Q-\al_2,\cdots) = r'(\al_1)-r'(\al_2)\ .
\end{align}
This equation allows us to check that $f_{12}$ is compatible with reflection symmetry, and to determine $r(\al)$ up to a linear function of $\al$. We can then compute $\pp{\al_1}F$ thanks to 
\begin{align}
 2\pp{\al_1}F = f_{12}(\al_1,\al_2,\cdots)-f_{12}(Q-\al_1,\al_2,\cdots)+ r'(\al_1)\ .
\end{align}
This determines $F$ up to a linear function of $\al_1$. By permutation symmetry, the ambiguity in $F$ amounts to the choice of two constants, which correspond to the parameters $\lambda$ and $\mu$ in eq. (\ref{zozz}).

\section{Peripheral topics}

\subsection{Higher Seiberg-Witten equations \label{apphsw}}

Here we argue that the following higher Seiberg-Witten equation holds for any $m\geq 1$:
\begin{align}
 \left(\pp{\al_j}-\pp{\al_i}\right)W_m(I) = 2\int_{z_i}^{z_j} \left(W_{m+1}(I,x)-\delta_{m,1}\frac{\frac12}{(x-x_1)^2}\right)dx\ ,
\label{hsw}
\end{align}
where $I=(x_1,\cdots x_m)$. These equations are generalizations of the Seiberg-Witten equation for $F$ (\ref{swp}), if we formally define $W_0$ by $W_0=F$. In this sense, the existence of higher Seiberg-Witten equations, which we will prove in the case $n=3$, lends support to the validity of the Seiberg-Witten equation for $F$. In addition, the higher Seiberg-Witten equations can be used for checking that $f_{ij}$ (\ref{fij}) obeys $\pp{\alpha_1}f_{23} + \pp{\alpha_2} f_{31}+\pp{\alpha_3}f_{12}=0$, which is necessary for $F$ to exist. 

There is however a substantial difference between the Seiberg-Witten equations for $F$ and for $W_m$. The former equation involves a specific choice of a combination of solutions $W_1(\overset{i+}{x})+W_1(\overset{j+}{x})$, in other words a specific determination of the integration variable $x$. The latter equations are however insensitive to this choice, as is indicated in eq. (\ref{hsw}) by the absence of superscripts for the variable $x$. 

To prove the higher Seiberg-Witten equations (\ref{hsw}), it is enough to show that they are compatible with the Ward identities (\ref{qpw}). This is because the correlation functions $W_m$ are unique solutions of these identities, as shown perturbatively in Section \ref{subsecewi}. The compatibility is easy to prove (not forgetting that $D_x$ in eq. (\ref{qpw}) acts on the integration bounds in eq. (\ref{hsw})), except
in the case of the $m=0$ Ward identity, which involves the function $t(x)$. Let us apply the differential operator $\pp{\al_j}-\pp{\al_i}$ to the $m=0$ Ward identity eq. (\ref{wiz}), assuming the higher Seiberg-Witten equations hold. The integrand in the resulting equation can be simplified with the help of the $m=1$ Ward identity, and we are left with the equation
\begin{align}
 \left(\pp{\al_j}-\pp{\al_i}\right)t(x) = \frac{Q}{(x-z_j)^2} -\frac{Q}{(x-z_i)^2} + 2\int_{z_i}^{z_j} \left(D_xW_1(y) +\pp{y}\frac{1}{x-y}W_1(y)\right)dy\ .
\end{align}
To prove this equation for $n\geq 4$ would require some information on the quantum accessory parameters which appear in $t(x)$ (\ref{tx}), and this is outside the scope of this article. However, in the case $n=3$, the accessory parameters 
are known. In this case, using the explicit expression (\ref{tf}) for $t(x)$ and the formula (\ref{dxw}) for $D_xW_1(y)$, the above equation is found to hold, provided 
\begin{align}
\forall k\ , \quad W_1(y) \underset{y\sim z_k}{=} \frac{-\al_k}{y-z_k} + o\left(\frac{1}{y-z_k}\right)\ .
\end{align}
This condition is obeyed by almost all solutions $W_1(y)$, provided $\Re(\al_k-\frac{Q}{2})<0$. This is easily understood in the heavy asymptotic limit, where our condition on $\al_k$ becomes $\Re \eta_k<\frac12$. Then all solutions $\Psi(y)$ of the linear differential equation (\ref{ptp}) behave as $\Psi(y) \underset{y\sim z_k}{=} O((y-z_k)^{\eta_k})$, except the particular solution $\Psi(\overset{k-}{y})$ which corresponds to the critical exponent $1-\eta_k$. Thus we have $W_1^{(0)}(y) =-\frac{\Psi'(y)}{\Psi(y)}\underset{y\sim z_k}{=} \frac{-\eta_k}{y-z_k} + o\left(\frac{1}{y-z_k}\right)$ for almost all solutions.

Our higher Seiberg-Witten equations are special cases of equations which describe the behaviour of $W_m$ under arbitrary deformations of the spectral curve, and which can be found in \cite{cem09} (equation (6.48) therein). 

\subsection{Comparison with matrix models \label{appmm}}

An $n$-point function $\la \prod_{i=1}^n V_{\al_i}(z_i,\bar{z}_i)\ra$ of Liouville theory at central charge $c=1+6Q^2=1+6(b+b^{-1})^2$ can be expressed as an $N$-dimensional integral provided \cite{zz95}
\begin{align}
 \sum_{i=1}^n \al_i = Q- Nb\ ,
\label{san}
\end{align}
where $N$ is a natural integer.
In this case, the functional integral description of Liouville theory implies that $\la \prod_{i=1}^n V_{\al_i}(z_i,\bar{z}_i)\ra$ has a pole, whose residue is (up to unimportant factors)
\begin{align}
 Z_N = \prod_{i<j} |z_{ij}|^{-4\al_i\al_j} \int dY \prod_{a=1}^N\prod_{i=1}^n |y_a-z_i|^{-4b^{-1}\al_i}\ .
\end{align}
where we introduce the following measure on the $N$-uple of complex variables $Y=(y_1,\cdots y_N)$
\begin{align}
 dY = \prod_{a< b}|y_{ab}|^{-4b^{-2}} \prod_{a=1}^N d^2y_a \ .
\end{align}
We can interpret $\{y_a\}$ as the eigenvalues of a complex matrix $M$ of size $N$, and $dY$ as a measure on the space of such matrices (the "beta ensemble"). This is the basis of the proof of the AGT relation by Dijkgraaf and Vafa in this case \cite{dv09}. 
The free energy $F_N=\log Z_N$ is moreover known to obey Seiberg-Witten equations. This was proved perturbatively in \cite{che10}, elaborated upon in \cite{bou12}, and checked in the related case of the Gaussian potential in \cite{mmps11}. 

It is possible to apply our formalism in this case, and the spin one chiral field is then 
\begin{align}
 J(x) = \Tr \frac{1}{x-M} = \sum_{a=1}^N \frac{1}{x-y_a}\ .
\end{align}
The hypergeometric solution $\Psi(\overset{1}{x})$ (\ref{pox}) of our second-order equation reduces to a Jacobi polynomial of degree $N$,
\begin{align}
 \Psi(\overset{1}{x}) = \prod_{i=1}^3 (x-z_i)^{\eta_i} \prod_{a=1}^N (x-s_a)\ ,
\end{align}
whose zeroes $s_a$ must obey 
\begin{align}
 \sum_{b\neq a}\frac{1}{s_a-s_b}+\sum_{i=1}^3 \frac{\eta_i}{s_a-z_i} = 0\ .
\label{sai}
\end{align}
Actually, the solution $\Psi(\overset{1}{x})$ has a diagonal monodromy around not only $z_1$, but also $z_2$ and $z_3$, and we have $\Psi(\overset{1}{x})=\Psi(\overset{2}{x})=\Psi(\overset{3}{x})$. It is natural to choose this solution in computing the correlation functions $W_m^{(g)}$, which will then be meromorphic functions. On the other hand, the subdominant solutions $\Psi(\overset{1-}{x})$, $\Psi(\overset{2-}{x})$ and $\Psi(\overset{3-}{x})$ remain different from each other, and their monodromy properties do not simplify.

So, in the special case (\ref{san}), the spin one chiral field $J$ is meromorphic, as in the free boson theory (which is the case $N=0$). It may seem that having polynomial solutions of the second-order equation brings important simplifications. And indeed, relatively simple formulas for the correlation functions $W_m^{(g)}$ can be obtained. However, these formulas depend not only on the physical variables $\eta_i,z_i$, but also on the auxiliary variables $s_a$. And it is not easy to eliminate the auxiliary variables in the resulting expression for the free energy $F^{(g)}$.
Therefore, when it comes to computing the free energy, the special case is not necessarily easier than the general case.

\acknowledgments{
We are grateful to Michel Berg\`ere for collaboration on related topics. We thank Ivan Kostov, Vincent Pasquier and especially Michel Berg\`ere for helpful comments on the manuscript. L. Chekhov is grateful to the Russian Foundation for Basic Research for support (Grants Nos. 11-01-00440-a, 11-01-12037-ofi-m-2011, and 11-02-90453-Ukr-f-a), to the Grant for Supporting Leading Scientific Schools
NSh-4612.2012.1, and to the Program Mathematical Methods of Nonlinear Dynamics.
}


\end{document}